\documentclass[letterpaper, 10 pt, conference]{ieeeconf}  

\IEEEoverridecommandlockouts                              

\usepackage{comment}
\usepackage[backend=biber,style=ieee,maxbibnames=6]{biblatex}
\usepackage{amsmath,amssymb,amsfonts}
\usepackage{tabularx}
\PassOptionsToPackage{dvipsnames,svgnames,table}{xcolor}
\usepackage{multirow}
\usepackage{graphicx}
\usepackage{textcomp}
\usepackage{xcolor}
\usepackage{array}
\usepackage{booktabs}
\usepackage[colorlinks=true,allcolors=blue]{hyperref}
\usepackage{placeins}
\usepackage{stfloats}
\usepackage{algorithm}
\usepackage{algpseudocode}
\usepackage{capt-of}
\usepackage{makecell}
\usepackage{subcaption}
\graphicspath{{figures/}}

\renewcommand{\arraystretch}{0.95}

\begin{document}


\title{Model-Free Stealthy Attack on Vehicle Systems and Its Implications for ISO/SAE 21434 Risk Assessment}


\author{Qazi Mairaj ud din, and Qadeer Ahmed%
\thanks{The authors are with the Center for Automotive Research, The Ohio
State University, Columbus, OH, USA (e-mail: mairajuddin.1@osu.edu;
 ahmed.358@osu.edu). Implementation available at: \url{https://github.com/OSU-CAR-MSL/stealthy-attack-risk-assessment}}}

\maketitle

\begin{abstract}
This paper investigates a model-free stealthy adversary targeting in-vehicle network and examines the implications for automotive cybersecurity risk assessment. Conventional stealthy false-data-injection (FDI) attacks often assume privileged knowledge of system dynamics, estimator parameters, or detector statistics, potentially understating the feasibility of adversaries operating with limited system information. We develop a model-free FDI attack that uses a Vector Autoregressive model with Exogenous inputs (VARX), identified from passively recorded nominal measurements, to characterize a compromised channel and construct attack sequences within a data-calibrated residual envelope. The construction requires no vehicle dynamics model, estimator parameters, assumed noise distribution, or detector knowledge. In a closed-loop lane-keeping benchmark, the attack induces a lateral displacement of 1.25\,m, exceeding the specified 0.9\,m half-lane width, and maintains the vehicle outside the lane corridor for 9.21\,s. At a matched 1\,\% nominal false-alarm operating point, attack-period alarm rates remain below 2.4\,\% across five learning-based in-vehicle intrusion detection systems (IDSs), a $\chi^2$ residual monitor, and a liveness monitor, compared with 66--100\,\% for conventional attack baselines. The proposed adversary achieves comparable physical impact and detector evasion to a privileged model-based adversary while requiring 13 fewer attack-potential points, increasing its assessed feasibility from Low to High under the considered ISO/SAE 21434 Threat Analysis and Risk Assessment (TARA). We further demonstrate that conventional impact-and-feasibility-based risk scoring can assign identical risk values to attacks exhibiting substantially different detector responses, and introduce a detection-conditioned risk formulation to distinguish them. These findings highlight the importance of accounting for adversarial knowledge requirements and measured detectability when assessing cybersecurity risks in safety-critical vehicle systems.
\end{abstract}

\noindent\textbf{Keywords:}
Vehicle Cybersecurity, CAN bus, false data injection, stealthy attacks,
intrusion detection, ISO/SAE~21434.


\section{Introduction}
\label{sec:intro}

Road vehicles have become distributed computing platforms whose safety
functions are realized in software and carried over internal networks. A modern
vehicle executes lane keeping, adaptive cruise control and emergency braking on
electronic control units (ECUs) that exchange sensor measurements and actuator
commands over a Controller Area Network (CAN), which lack message
authentication or integrity protection due to higher computational cost and limited bandwidth. Experimental analyses have shown that a vehicle network can be reached remotely through attack surfaces such as Electronic Logging Devices (ELDs), telematics, infotainment and diagnostic interfaces~\cite{koscher2010experimental, jepson2024commercial}, and documented vehicle thefts involving inexpensive CAN-injection devices demonstrate that adversaries can obtain physical access to in-vehicle communication networks through externally accessible components~\cite{Tindell2023CANInjection}. Automotive cybersecurity risk management has also become an integral part of vehicle type-approval requirements. UN Regulation No.~155 requires manufacturers to demonstrate a compliant Cybersecurity Management System (CSMS) spanning vehicle development, production, and post-production, including the identification, assessment, and management of cybersecurity risks~\cite{unr155}. Complementing these regulatory requirements, ISO/SAE 21434 establishes a cybersecurity engineering framework that incorporates Threat Analysis and Risk Assessment (TARA)~\cite{iso21434}. Within the TARA framework, the resulting attack-feasibility rating is combined with the impact rating of the associated damage scenario to determine cybersecurity risk. However, these ratings do not explicitly quantify an attack's ability to evade a deployed IDS. Consequently, two attack scenarios assigned identical impact and feasibility ratings may receive the same risk value despite exhibiting substantially different detection rates against the vehicle's monitoring mechanisms. This distinction motivates evaluating adversarial capabilities not only in terms of the resources and knowledge required to execute an attack, but also in terms of its detectability and achievable physical impact.


\subsection{Related Work}
\label{subsec:related}

Three main directions in the existing literature address this question.

\emph{Model-based stealthy attacks.} Mo and Sinopoli characterized the
conditions under which an adversary that injects false sensor data holds the
estimator residual within its nominal distribution while the estimation error
grows without bound~\cite{mo2010fdia, mo2015performance}, and Pasqualetti
\emph{et al.} established the corresponding detectability limits for descriptor
models of cyber-physical systems~\cite{pasqualetti2013attack}. Teixeira
\emph{et al.} decomposed adversary capability into model knowledge, disclosure
resources and disruption resources, and showed how each named attack occupies a
distinct point in that space~\cite{teixeira2015secure}. Bai \emph{et al.} derived
performance bounds for Kalman filtering with compromised sensors under a
stealthiness constraint~\cite{bai2017kalman}. These model-based formulations provide rigorous characterizations of stealthy attack conditions and achievable estimation degradation, but their guarantees rely on substantial knowledge of the target system, including its dynamics, estimator parameters, and relevant noise statistics. Such assumptions impose a significant information requirement on the adversary. Under the attack-potential-based assessment of ISO/SAE 21434, obtaining this information is reflected in the Knowledge of the Item factor, potentially increasing the required attack potential and reducing the assessed attack feasibility~\cite{iso21434}. Consequently, although these formulations establish the theoretical possibility of stealthy attacks, their applicability to adversaries operating without privileged system knowledge remains limited. This limitation motivates a second line of research that seeks to construct stealthy attacks from observed system behavior rather than explicit knowledge of the underlying dynamics and detection mechanisms.

\emph{Learning-based attack generation.} Khazraei \emph{et al.} extended stealthy attack construction beyond conventional model-based formulations by introducing learning-enabled attack generators for nonlinear cyber-physical systems~\cite{khazraei2022learning}. Their grey-box framework employs feed-forward and recurrent neural networks trained offline to balance state-estimation degradation against deviations in the innovation residual used for anomaly detection. Although the recurrent architecture reduces the information required during deployment, training still relies on access to the estimator model and internal quantities, including state estimates and estimator gains. Consequently, the framework relaxes the requirement for explicit plant knowledge but does not eliminate dependence on privileged estimator information. Furthermore, the evaluation primarily considers estimator-residual-based anomaly detection rather than contemporary IDSs operating on in-vehicle network traffic. These limitations leave open whether impactful stealthy attacks can be learned entirely from passively recorded nominal measurements, without access to estimator internals or detector-specific information.

\emph{In-vehicle intrusion detection.} In parallel, substantial research has focused on detecting anomalous behavior in Controller Area Network (CAN) traffic using learning-based IDSs~\cite{rajapaksha2023ai}. CANet employs identifier-specific recurrent encoders and a shared latent representation to capture dependencies among CAN signals and detect anomalies through reconstruction errors~\cite{hanselmann2020canet}. Similarly, INDRA uses a gated recurrent unit (GRU)-based autoencoder to reconstruct signal sequences and identify intrusions through reconstruction-based anomaly scores~\cite{kukkala2020indra}. NeuroCAN incorporates contextual information from multiple message identifiers to capture inter-signal temporal dependencies~\cite{balaji2021neurocan}, while spatiotemporal deep-learning architectures further exploit relationships within vehicle-network traffic~\cite{mansourian2023deep}. 

These methods are commonly evaluated against conventional attack scenarios, including Denial-of-Service (DoS), fuzzing, spoofing, replay, plateau, and message suppression attacks, although the evaluated attack classes vary across studies~\cite{hanselmann2020canet,kukkala2020indra,song2020vehicle}. Such attacks often introduce detectable deviations in message frequency, payload values, or temporal and contextual relationships. However, strong detection performance against these conventional attack patterns does not necessarily establish robustness against stealthy attacks that are adversarially crafted to remain consistent with nominal network behavior while inducing significant deviations in the underlying physical system.

To address these gaps, this paper makes four contributions:

\begin{enumerate}
    \item We develop a data-driven stealthy False-Data Injection (FDI) framework for in-vehicle networks that learns the nominal behavior of a target measurement channel using a Vector Autoregressive model with Exogenous inputs (VARX). A residual envelope calibrated from passively recorded CAN traffic guides the construction of
    measurement-consistent attack sequences, without requiring knowledge of the vehicle dynamics, state estimator, assumed noise distribution, or detector parameters.

    \item We compare the proposed measurement-only adversary with a privileged model-based adversary possessing the plant dynamics, estimator gain, and innovation covariance. The proposed construction achieves comparable physical impact and detector evasion while reducing the assessed attack-potential score by 13 points. Under the considered ISO/SAE 21434 TARA assessment, this reduction changes the attack-feasibility rating from Low to High, demonstrating how reduced information requirements can materially alter the assessed feasibility of a stealthy attack.

    \item We conduct an ISO/SAE 21434-based TARA of the proposed attack and conventional attack baselines. Despite substantial differences in their observed detectability, the standard impact-and-feasibility-based risk calculation assigns the same maximum risk value to the proposed attack and four conventional   baselines. To expose this distinction, we introduce a detection-conditioned risk formulation that explicitly incorporates the measured response of deployed intrusion detection mechanisms, complementing the conventional TARA assessment.

    \item We validate the proposed attack on a closed-loop lane-keeping system using an unseen driving maneuver. The attack induces a maximum lateral displacement of 1.25\,m, exceeding the specified 0.9\,m half-lane width, and maintains the vehicle outside the lane corridor for 9.21\,s. We further evaluate its detectability using five learning-based in-vehicle IDSs, a $\chi^2$ innovation-residual monitor, and a liveness monitor. At a matched nominal false-alarm operating point of 1\,\%, none of these monitors exhibits an attack-period alarm rate exceeding 2.4\,\%, whereas the evaluated conventional attack baselines yield alarm rates ranging from 66\,\% to 100\,\% in the reported comparisons.
\end{enumerate}

\begin{figure}[t]
    \centering
    \includegraphics[width=\linewidth]{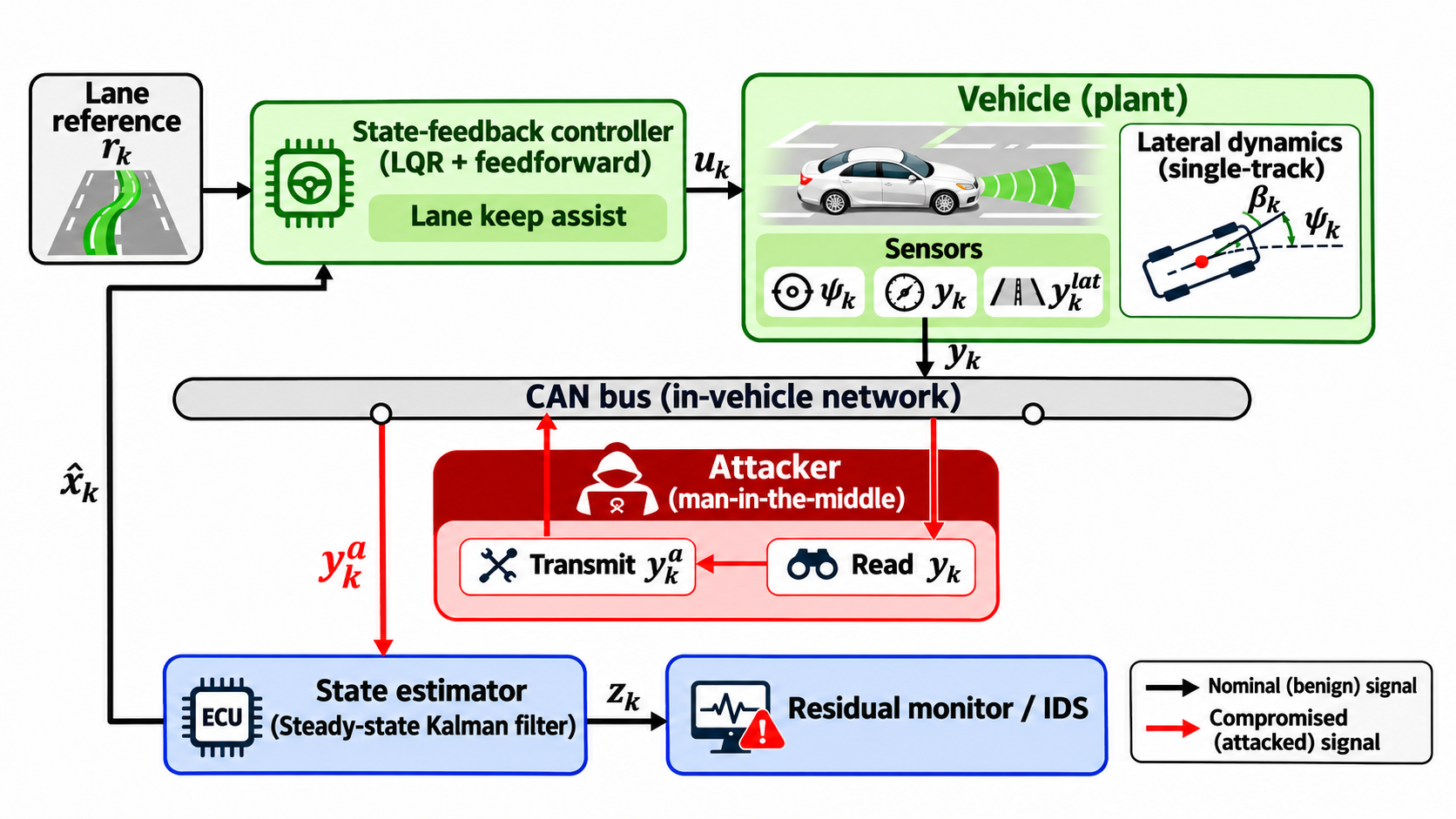}
    \caption{Vehicle system and proposed attack Model.}
    \label{fig:lateral_attack}
\end{figure}
Fig.~\ref{fig:lateral_attack} shows a typical vehicle setting with lateral control and our proposed attack model. The remainder of the paper is organized as follows.
Section~\ref{sec:model} states the system, the adversary and the attack
construction. Section~\ref{sec:results} reports the experimental setup, the
impact and detection results, and the risk assessment.
Section~\ref{sec:conclusion} concludes.


\section{System and Attack Model}
\label{sec:model}

\subsection{Plant, Estimator and Controller}
\label{subsec:plant}

The item under consideration is the lateral-control function of a passenger
vehicle, whose dynamics are represented by a single-track model and discretized
at a sample period $T_s$,
\begin{align}
  x_{k+1} &= A x_k + B u_k + w_k, & w_k &\sim \mathcal{N}(0,W), \label{eq:plant}\\
  y_k     &= C x_k + v_k,         & v_k &\sim \mathcal{N}(0,V), \label{eq:meas}
\end{align}
where $x_k$ is the state, $u_k$ the steering command, $y_k$ the transmitted
measurement, and $w_k$ and $v_k$ are mutually independent zero-mean Gaussian
process and measurement noises with covariances $W$ and $V$. The state matrix
$A=e^{A_cT_s}$ and the input matrix $B=\int_0^{T_s}e^{A_c\sigma}B_c\,\mathrm{d}\sigma$ are obtained from the
continuous-time single-track state and input matrices $A_c$ and $B_c$, and $C$
selects the measured states. The state
$x_k=[\beta_k,\ \dot{\psi}_k,\ \psi_k,\ y^{\mathrm{lat}}_k]^{\top}$ comprises the
sideslip angle, the yaw rate, the heading angle and the lateral offset from the
lane center, and the measurement
$y_k=[\dot{\psi}_k,\ \psi_k,\ y^{\mathrm{lat}}_k]^{\top}$ is transmitted as one
multi-signal message on the in-vehicle bus. The pair $(A,C)$ is observable, and
the lateral-offset state is reached through an integrating mode, which is the
structural property the attack exploits.

A steady-state Kalman filter forms the one-step prior
$\hat{x}_{k|k-1}=A\hat{x}_{k-1}+Bu_{k-1}$ from the previous filtered estimate
$\hat{x}_{k-1}$, and the innovation
\begin{equation}
  z_k = y_k - C\hat{x}_{k|k-1} ,
  \label{eq:innov}
\end{equation}
whose steady-state covariance is denoted $S$. A discrete-time LQR regulator with
feedforward gain closes the loop on the estimate,
$u_k=-L\hat{x}_k+N_{\mathrm{ff}}r_k$, where $L$ is the regulator gain,
$N_{\mathrm{ff}}$ the feedforward gain and $r_k$ the commanded lateral
reference. A residual monitor thresholds the normalized statistic
$g_k=z_k^{\top}S^{-1}z_k$ at its $(1-\alpha)$ nominal quantile with
$\alpha=1\%$. Because the controller acts on $\hat{x}_k$, any bias the filter
accepts into the lateral-offset estimate is transferred to the true lateral
position by the feedback law itself.

\subsection{Adversary Model}
\label{subsec:adversary}

The adversary's capabilities are consolidated into three assumptions that hold
for legacy and remotely accessible in-vehicle networks. Each one is also the
origin of an attack-potential factor rated in Section~\ref{subsec:risk}.

\emph{Frame access.} The in-vehicle network, CAN bus, carries no message authentication or integrity protection, so a node that reaches the bus reads every frame and overwrites the frames it selects~\cite{jepson2024commercial,Tindell2023CANInjection}.
Payloads are decoded into physical signals from public specifications such as
SAE~J1939 or from reverse-engineered signal databases (e.g.\ opendbc). The tooling is a laptop and a CAN interface, which is the Equipment factor at its lowest value.

\emph{Passive observation.} The adversary records long, high-quality traffic of
attack-free operation without being detected~\cite{7467419}. Recording is the
only preparation the attack requires, and it sets the Elapsed Time and Window of
Opportunity factors.

\emph{Operating envelope.} The operating distribution at deployment matches that
of the nominal recording, which is the condition under which an identified model
transfers from the recording to the attacked run.

In a realistic setting, the attack is resource-constrained~\cite{guo2022stealthy}: the adversary modifies a single channel. The \emph{target channel} carries the lateral offset $y^{\mathrm{lat}}_k$ and is the one the adversary writes, from a man-in-the-middle position on the path between the lateral-offset sensor and the lateral-control ECU. The \emph{context channels} carry the yaw rate $\dot{\psi}_k$ and the heading $\psi_k$; the adversary reads them, together with the steering command $u_k$, and transmits them unmodified. In the resource decomposition of Teixeira~\emph{et al.}~\cite{teixeira2015secure}, the disclosure resource is bus traffic already present on the network, and the disruption resource is one signal of one message.

The third resource in that decomposition is model knowledge, and the model-free adversary has none of it. The attack requires no plant model, no sensing map, no estimator or detector parameters: none of $A$, $B$, $C$, the Kalman gain, the innovation covariance $S$, the regulator gain $L$, the noise laws $W$ and $V$, the detector structure or the detector threshold enters the construction. Every quantity it uses is recovered from the passive measurement recording. This
is the Knowledge of the Item factor, and it is the factor the present
contribution moves, from confidential design data to public information. The
conventional baseline attacks of Section~\ref{subsec:baselines} also occupy that
position on that factor; the model-based stealthy reference is the only attack
evaluated here that requires more privileged information, namely the item's design data.

\subsection{Model-Free Identification of the Target Channel}
\label{subsec:varx}

In the model-free approach, the adversary fits a Vector Auto-Regressive model with Exogenous inputs (VARX) to the recorded target channel, taking the steering command and the context yaw rate as exogenous regressors:
\begin{equation}
  y^{\mathrm{lat}}_k
   = \sum_{i=1}^{p} a_i\, y^{\mathrm{lat}}_{k-i}
   + \sum_{i=1}^{p} b_i^{\top}\,\xi_{k-i}
   + c + e_k ,
  \qquad
  \xi_k = \begin{bmatrix} u_k \\ \dot{\psi}_k \end{bmatrix},
  \label{eq:varx}
\end{equation}
where $p$ is the model order, $a_i$ are the scalar autoregressive coefficients
of the target channel, $b_i$ are the coefficient vectors of the exogenous
regressors $\xi_k$, $c$ is a constant offset, and $e_k$ is the one-step
prediction error. The coefficients are obtained by regularized least squares on
the recorded data. Equation~\eqref{eq:varx} is the innovations representation of
the target channel: for a linear plant under a steady-state Kalman filter, the
one-step prediction error of a sufficiently long VARX fit converges to the
Kalman innovation on that channel~\cite{lutkepohl2005new,ljung1999system}, so the
adversary recovers the quantity a residual monitor thresholds without recovering
the model that generates it.

The adversary then calibrates its own stealth envelope on the same recorded
data,
\begin{equation}
  \tau = Q_{1-\alpha}\big(|e_k|\big),
  \label{eq:tau}
\end{equation}
where $Q_{1-\alpha}(\cdot)$ denotes the empirical $(1-\alpha)$ quantile of its
argument over the recording, so $\tau$ is the magnitude the absolute one-step
prediction error exceeds on a fraction $\alpha$ of nominal samples. The envelope
is an estimate the adversary forms from traffic; it is not the defender's
threshold, which the adversary never observes.

\subsection{Injection Construction}
\label{subsec:injection}

From the onset instant $k_0$ the adversary transmits
\begin{equation}
  \tilde{y}^{\mathrm{lat}}_k
    = \hat{y}^{\mathrm{lat}}_{k\mid k-1}
    + s\,\beta\tau\,\min\!\Big(1,\tfrac{k-k_0}{n_{\mathrm{ramp}}}\Big),
  \qquad s\in\{+1,-1\},
  \label{eq:inject}
\end{equation}
where $\hat{y}^{\mathrm{lat}}_{k\mid k-1}$ is the one-step VARX prediction
of~\eqref{eq:varx}, $\beta$ is the envelope multiple that sets the operating
point, $n_{\mathrm{ramp}}$ is the ramp length and $s$ selects the direction of
the departure. The predictor is advanced on the \emph{true} measurement the
man-in-the-middle observes rather than on the value it transmits, which keeps
the construction bounded on the integrating lateral mode.

The attack mechanism is the following. The transmitted attacked value tracks the predicted nominal evolution of the channel and the vehicle dynamics, so the increment between consecutive transmitted samples remains consistent with the dynamics the detectors learned. The persistent component $\beta\tau$ is applied gradually, so the Kalman filter absorbs it into $\hat{x}_k$ instead of expressing it in $z_k$. The regulator acts on the biased estimate and steers the vehicle until the \emph{true} lateral offset carries the bias that the \emph{reported} offset does not. Impact is produced by the controller's action in the loop and not by the magnitude of the injection.

\subsection{Stealth Criterion}
\label{subsec:stealth}

An attack matters to a vehicle cybersecurity assessment only when it is both
consequential and undetectable, and the two conditions are evaluated at the
operating point at which the monitors are actually deployed. Let
$\hat{A}_j$ denote the alarm rate of monitor $j$ measured over the attack
window, with every monitor in the deployed set $\mathcal{M}$ thresholded at its
own nominal false-alarm rate. The attack is successful when
\begin{equation}
  \underbrace{\max_{j\in\mathcal{M}}\ \hat{A}_j \;\le\; \epsilon}_{\text{stealth}}
  \qquad\text{and}\qquad
  \underbrace{\max_{k\ge k_0}\ \big|y^{\mathrm{lat}}_k - r_k\big|
    \;\ge\; \delta_{\mathrm{lim}}}_{\text{hazard}} ,
  \label{eq:eps-alpha}
\end{equation}
that is, when no deployed monitor reports an alarm rate above the nominal
false-alarm level $\epsilon$ while the vehicle leaves the safe lane-keeping
limit $\delta_{\mathrm{lim}}$. Both terms are measured quantities rather than
modeling assumptions: the left term is read from the detector battery at a
matched false-alarm operating point and the right term from the vehicle
trajectory. Section~\ref{subsec:setup} fixes $\epsilon=0.03$, the $1\%$ nominal
false-alarm rate with a two-percentage-point margin, and
$\delta_{\mathrm{lim}}=0.9$\,m, the half-lane width.

\subsection{Baseline Attacks}
\label{subsec:baselines}

Six baselines span the capability positions of Section~\ref{subsec:adversary}.
\emph{Denial of service} withholds the target channel measurement, so the estimator runs predict-only on that channel; it is the signal-level counterpart of the
message-suppression and bus-flooding attacks used to evaluate in-vehicle
detectors~\cite{hanselmann2020canet,song2020vehicle}. \emph{Fuzzing} transmits
uniformly random values drawn from the observed range of the channel, which is
the signal-level counterpart of the random-payload fuzzy attacks in the public
CAN intrusion datasets~\cite{song2020vehicle}. \emph{Replay} retransmits a recorded
nominal segment at a fixed lag, the playback attack of~\cite{hanselmann2020canet,kukkala2020indra}. \emph{Plateau} holds the channel at its value at the onset instant, the plateau attack of the same
references~\cite{hanselmann2020canet,kukkala2020indra}. \emph{Gaussian injection} adds
zero-mean noise of a fixed standard deviation to the true value, representing a
degraded or noisy sensor rather than a deliberate manipulation. The model-based
stealthy baseline is the privileged reference: it replaces the VARX predictor
of~\eqref{eq:inject} with the system model, the estimator and the detector
statistics, and is otherwise identical. This baseline isolates what the removal
of model knowledge costs.

\section{Experimental Setup and Results}
\label{sec:results}

\subsection{Setup}
\label{subsec:setup}

\emph{Scenario.} The plant of Section~\ref{subsec:plant} is instantiated on a
benchmark electric-vehicle model~\cite{rana2020iot} with the parameters of
Table~\ref{tab:veh}, discretized at $T_s=1$\,ms. Nominal traffic is generated
on two maneuvers, a sinusoidal tracking profile and a double lane change, and is
the only data available to the adversary and to the detectors. The attack is
mounted on a third, held-out maneuver, a single lane change, so neither the
adversary's identified model nor the defender's detectors have been exposed to
the maneuver under attack. The deployment horizon is $12000$ samples, that is
12\,s, and the injection starts at $k_0=2400$, that is
2.4\,s, leaving 9.6\,s of attacked operation. Every result is
the mean over $3$ deployment seeds, so that no reported quantity depends
on a single noise realization.


\begin{table}[t]
  \centering
  \caption{Benchmark vehicle parameters~\cite{rana2020iot}.}
  \label{tab:veh}
  \setlength{\tabcolsep}{10pt}\footnotesize
  \begin{tabular}{llll}
    \toprule
    Item & Value & Item & Value\\
    \midrule
    $T_s$   & 1\,ms            & $M$     & 380\,kg\\
    $l_r$   & 0.6\,m             & $l_f$   & 0.8\,m\\
    $C_r$   & 6000\,N/rad        & $C_f$   & 6000\,N/rad\\
    $V_x$   & 5.5556\,m/s        & $l_{\mathrm{pre}}$ & 5.135\,m\\
    $J$     & 136.08\,N$\cdot$m/(rad/s$^2$) & half-lane & 0.9\,m\\
    \bottomrule
  \end{tabular}
\end{table}

\emph{Adversary configuration.} The VARX order is $p=40$. Two measurements
establish that the fit of~\eqref{eq:varx} recovers the innovation it is intended
to recover. On a reference linear system whose Kalman innovation is known in
closed form, the VARX one-step prediction error correlates with the true
innovation at $0.99925$ and reproduces its variance to within
$0.34\,\%$ ($0.07739$ against $0.07713$). On the vehicle itself the
learned envelope is $\tau=0.1388$ against a true Kalman innovation envelope of
$0.1375$ on the same channel, a difference of $0.91\,\%$. The adversary
therefore reconstructs, from recorded traffic alone, the statistic a
residual-based monitor thresholds. 

\emph{Detectors.} Five learned in-vehicle intrusion detection systems are
trained on nominal traffic of the two training maneuvers only, so no detector
observes the maneuver under attack. CANet is an LSTM encoder with a joint latent
bottleneck and a fully connected decoder that reconstructs the current
multi-signal message~\cite{hanselmann2020canet}. INDRA is a recurrent
autoencoder, a linear embedding followed by GRU encoder and decoder stages and a
linear output layer, that reconstructs the
subsequence~\cite{kukkala2020indra}. NeuroCAN~\cite{balaji2021neurocan} and the
two deep spatiotemporal detectors of~\cite{mansourian2023deep}, denoted DL-LSTM
and DL-CNNLSTM, complete the battery. All five ingest the three signals of the
message jointly and are scored by the message intrusion score, the maximum over
signals of the per-signal reconstruction squared error, so a manipulation of one
signal is detectable through the broken dependency with the context channels.
Two further monitors are reported: the $\chi^2$ residual monitor on $g_k$~\cite{mo2015performance}, the classical
fault-diagnosis test, which operates in the residual domain, and a liveness monitor, which reports the absence of the message. Every monitor is thresholded at the $(1-\alpha)$ quantile of its own nominal score with $\alpha=1\%$, so all alarm rates are read at a matched false-alarm operating point.

\emph{Metrics.} Three quantities are reported here. Impact is the peak departure from the $\pm0.9$\,m lane corridor in meters and the dwell time outside it in seconds; severity follows the lane-keeping requirement, rated S0 when the corridor is never exited, S2 for an excursion shorter than 1\,s and S3 for a sustained departure. Stealth is the alarm rate of each monitor at the matched false-alarm operating point; the stealth condition of~\eqref{eq:eps-alpha} is evaluated at $\epsilon=0.03$, the $1\%$ nominal false-alarm rate with a two-percentage-point margin, and the hazard condition at $\delta_{\mathrm{lim}}=0.9$\,m. Cost is the effort the attack demands of the adversary, rated on the five attack-potential factors of ISO/SAE 21434 in Section~\ref{subsec:risk}. Impact and stealth are measurements and cost is a rating, so the three are reported separately; Section~\ref{subsec:risk} is where they are combined, and the combination is the risk value the standard defines.

\subsection{Impact and Attack Detection}
\label{subsec:impact}

\begin{table}[t]
\centering
\caption{Vehicle-behavior impact during a lane-change maneuver.}
\label{tab:impact}

\footnotesize
\setlength{\tabcolsep}{3pt}
\renewcommand{\arraystretch}{1.08}

\begin{tabular}{@{}lcrrc@{}}
\toprule
Attack
& \shortstack{Model\\knowledge}
& \shortstack{Peak\\dep. (m)}
& \shortstack{Dwell\\(s)}
& Severity \\
\midrule
Model-based stealthy
    & Full & 1.25 & 9.20 & S3 \\
\textbf{Model-free stealthy (ours)}
    & None & 1.27 & 9.22 & S3 \\
DoS
    & None & 1.02 & 1.02 & S3 \\
Fuzzing
    & None & 24.88 & 9.01 & S3 \\
Replay
    & None & 23.17 & 6.36 & S3 \\
Plateau
    & None & 63.95 & 8.96 & S3 \\
Gaussian
    & None & 0.16 & 0.00 & S0 \\
\bottomrule
\end{tabular}
\end{table}

\begin{figure*}[!t]
    \centering

    \begin{subfigure}[t]{0.32\textwidth}
        \centering
        \includegraphics[width=\linewidth]
        {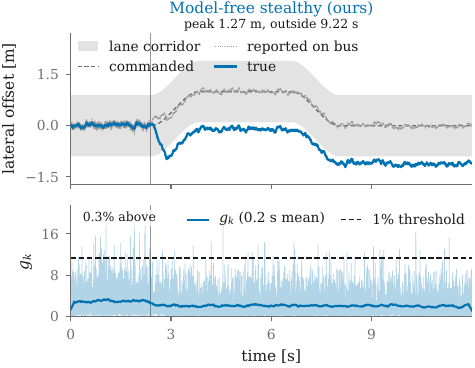}
        \caption{Model-free stealthy}
        \label{fig:attack-mf}
    \end{subfigure}
    \hfill
    \begin{subfigure}[t]{0.32\textwidth}
        \centering
        \includegraphics[width=\linewidth]
        {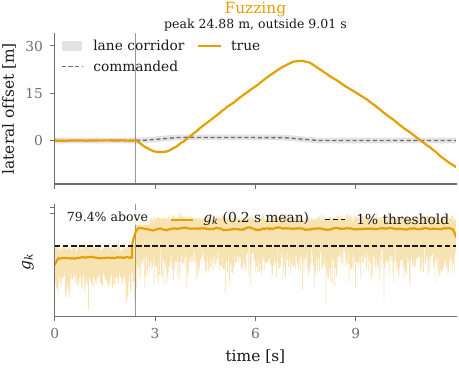}
        \caption{Fuzzing}
        \label{fig:attack-fuzz}
    \end{subfigure}
    \hfill
    \begin{subfigure}[t]{0.32\textwidth}
        \centering
        \includegraphics[width=\linewidth]
        {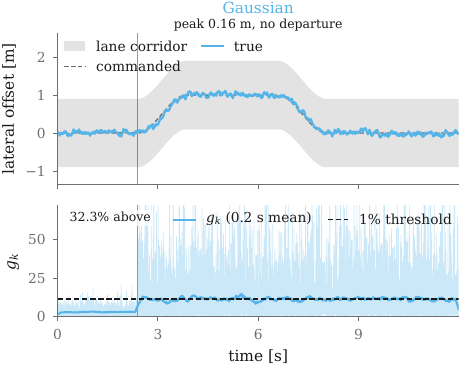}
        \caption{Gaussian injection}
        \label{fig:attack-gauss}
    \end{subfigure}

    \caption{Lateral offset and residual statistic $g_k$ during the lane-change maneuver. The vertical line indicates attack onset at $2.4$\,s; the dashed line denotes the nominal $1\%$ false-alarm threshold. Additional results are reported in Tables~\ref{tab:impact} and~\ref{tab:ids}.}
    \label{fig:attacks}
\end{figure*}

Table~\ref{tab:impact} reports the impact metrics and the model knowledge each
attack requires, and Fig.~\ref{fig:attacks} shows the closed-loop behavior of
three representative attacks together with the residual statistic.

The model-free stealthy attack drives the vehicle 1.25\,m from the commanded
lateral reference, against a half-lane width of 0.9\,m, and holds it
outside the lane corridor for 9.21\,s of the 9.6\,s of attacked
operation. The reported lateral offset meanwhile follows the commanded lane
change within the corridor. Fig.~\ref{fig:attacks}(a) makes that separation
visible: the value on the bus and the true position of the vehicle diverge by
roughly the injected offset and remain divergent for the rest of the run. The
conventional baselines produce larger displacement, up to 63.95\,m for
plateau, which is the expected outcome of an unconstrained manipulation and is
the reason displacement alone does not order these threats.

\begin{table*}[t]
\centering
\caption{Alarm rates of intrusion detection monitors at the
1\% nominal false-alarm operating point.}
\label{tab:ids}

\footnotesize
\setlength{\tabcolsep}{4pt}
\renewcommand{\arraystretch}{1.05}

\begin{tabular}{@{}lrrrrrrr@{}}
\toprule
& \multicolumn{2}{c}{Stealthy attacks}
& \multicolumn{5}{c}{Conventional attacks} \\
\cmidrule(lr){2-3}
\cmidrule(lr){4-8}

Monitor
& \shortstack{\textbf{Model-free}\\\textbf{(ours)}}
& \shortstack{Model-\\based}
& DoS
& Fuzzing
& Replay
& Plateau
& Gaussian \\
\midrule

CANet~\cite{hanselmann2020canet}
& \textbf{0.021} & 0.020 & 0.000 & 0.999
& 0.776 & 0.653 & 0.044 \\

INDRA~\cite{kukkala2020indra}
& \textbf{0.016} & 0.015 & 0.000 & 0.987
& 0.654 & 0.611 & 0.215 \\

NeuroCAN~\cite{balaji2021neurocan}
& \textbf{0.022} & 0.021 & 0.000 & 0.999
& 0.777 & 0.653 & 0.113 \\

DL-LSTM~\cite{mansourian2023deep}
& \textbf{0.016} & 0.015 & 0.000 & 0.998
& 0.791 & 0.656 & 0.039 \\

DL-CNNLSTM~\cite{mansourian2023deep}
& \textbf{0.024} & 0.022 & 0.000 & 0.998
& 0.792 & 0.662 & 0.059 \\

\midrule

$\chi^2$ residual monitor
& \textbf{0.003} & 0.003 & 0.003 & 0.792
& 0.080 & 0.046 & 0.324 \\

Liveness monitor
& \textbf{0.000} & 0.000 & 1.000 & 0.000
& 0.000 & 0.000 & 0.000 \\

\bottomrule
\end{tabular}
\end{table*}

Table~\ref{tab:ids} reports the attack detection result. Against the model-free attack, CANet alarms on 2.1\,\% of samples and INDRA on 1.6\,\%,
against the $1\%$ rate each produces on nominal traffic. No monitor in the
battery exceeds 2.4\,\% on the pooled mean. The $\chi^2$
residual monitor alarms on 0.3\,\% of samples, below its own nominal
false-alarm rate, which is the signature of an injection the filter has absorbed
into its state estimate rather than expressed in the innovation.

The baselines separate sharply. CANet and INDRA alarm on 99.9\,\%
and 98.7\,\% of fuzzing samples, on 77.6\,\% and
65.4\,\% of replay samples, and on 65.3\,\% and
61.1\,\% of plateau samples. Denial of service is invisible to
all five reconstruction detectors, because the target channel carries no data
for them to reconstruct, and is detected with certainty by the liveness monitor
at zero delay. 
Gaussian injection perturbs the channel at the operating point tested but never
drives the vehicle out of the lane corridor, with a peak departure of
0.15\,m, and is therefore not an impact-bearing attack irrespective of its
alarm rate.


\subsection{Cost of Removing Model Knowledge}
\label{subsec:cost}

The privileged model-based construction attains a peak departure of 1.25\,m
and a dwell time of 9.20\,s, against 1.25\,m and 9.21\,s for the
model-free construction, with a peak alarm rate of 2.2\,\% against
2.4\,\%. The two differ by 0.37\,\% in peak departure, and
the ordering is the same in each of the three seeds. Removing $A$, $C$, the
Kalman gain, the innovation covariance and the noise laws therefore costs no
measurable impact and no measurable stealth on this item.

The selected operating point is the largest injection compatible with the
stealth condition of~\eqref{eq:eps-alpha} rather than the largest injection
available. Peak departure grows affinely with the envelope multiple, reaching
4.60\,m at $\beta=32$, so an adversary that accepts detection can
obtain more displacement; $\beta=8$ is the largest multiple at which
CANet and INDRA remain at their nominal alarm rates.

\subsection{Attack Feasibility and Risk}
\label{subsec:risk}

ISO/SAE 21434~\cite{iso21434} rates attack feasibility by the attack-potential
approach of ISO/IEC 18045~\cite{iso18045}, summing scores on Elapsed Time (ET),
Specialist Expertise (SE), Knowledge of the Item or Component (KoIC), Window of
Opportunity (WoO) and Equipment (Eq). Attack potential measures the effort an
adversary must expend and attack feasibility measures the ease of mounting the
attack, so the two run in opposite directions: an adversary economical on every
factor has minimal required attack potential and maximal attack feasibility.

\begin{table}[t]
\centering
\caption{Attack potential and feasibility ratings based on
ISO/IEC 18045 and ISO/SAE 21434.}
\label{tab:ap}

\footnotesize
\setlength{\tabcolsep}{2.5pt}
\renewcommand{\arraystretch}{1.08}

\begin{tabular}{@{}lrrrrrrl@{}}
\toprule
Attack & ET & SE & KoIC & WoO & Eq & Sum & Rating \\
\midrule
\shortstack[l]{\textbf{Model-free}\\\textbf{stealthy (ours)}}
    & 1 & 3 & 0 & 4 & 0 & \textbf{8} & High \\

\shortstack[l]{Model-based\\stealthy}
    & 4 & 6 & 7 & 4 & 0 & 21 & Low \\

DoS
    & 0 & 0 & 0 & 1 & 0 & 1 & High \\

Fuzzing
    & 0 & 0 & 0 & 1 & 0 & 1 & High \\

Replay
    & 1 & 3 & 0 & 4 & 0 & 8 & High \\

Plateau
    & 0 & 0 & 0 & 1 & 0 & 1 & High \\

Gaussian
    & 0 & 0 & 0 & 1 & 0 & 1 & High \\
\bottomrule
\end{tabular}

\vspace{2pt}
\parbox{\columnwidth}{%
\scriptsize
\textit{Note:} Lower scores indicate higher attack feasibility.
}
\end{table}

Table~\ref{tab:ap} assigns the values of ISO/SAE 21434 Table~G.6 to the
capabilities set out in Section~\ref{subsec:adversary}. The model-free adversary
requires elapsed time of at most one week, because identification is a
least-squares fit on recorded traffic and no vulnerability discovery is
involved; proficient rather than expert specialist expertise, because the
procedure is standard time-series identification and not a vehicle-specific
skill; public knowledge of the item, because no parameter of the plant, the
estimator, the controller or the detector enters the construction; a moderate
window of opportunity, because the adversary must both record nominal traffic
and hold a write position on one channel; and standard equipment, a laptop and a
CAN interface. The sum is 8, which maps to a High attack feasibility
rating under Table~G.7. The model-based adversary requires elapsed time of at
most one month, expert expertise, and confidential knowledge of the item, since
$A$, $C$, the Kalman gain and the innovation covariance are design data. Its sum
is 21, which maps to Low. Removing the knowledge requirement reduces
the required attack potential by 13 points and raises the feasibility
rating by two bands, while Section~\ref{subsec:cost} establishes that the impact
rating is unchanged.

The impact rating follows the lane-keeping requirement. A sustained departure
from the $\pm0.9$\,m corridor at road speed is an unintended lane
departure and is rated Severe; an excursion shorter than one second is rated
Major; a run that never exits the corridor is rated Negligible.

\begin{table}[t]
\centering
\caption{Attack risk assessment with proposed detection-aware adjustment.}
\label{tab:risk}

\footnotesize
\setlength{\tabcolsep}{3pt}
\renewcommand{\arraystretch}{1.08}

\begin{tabular}{@{}llcrcrr@{}}
\toprule
Attack & Impact & Feas. & $R$ & Det. & $E$ & $R_{\det}$ \\
\midrule
\shortstack[l]{\textbf{Model-free}\\\textbf{stealthy (ours)}}
    & Severe & High & 5 & No & 0.98 & \textbf{4.91} \\

\shortstack[l]{Model-based\\stealthy}
    & Severe & Low & 3 & No & 0.98 & 2.96 \\

DoS
    & Severe & High & 5 & Yes & 0.00 & 1.00 \\

Fuzzing
    & Severe & High & 5 & Yes & 0.00 & 1.01 \\

Replay
    & Severe & High & 5 & Yes & 0.21 & 1.83 \\

Plateau
    & Severe & High & 5 & Yes & 0.34 & 2.35 \\

Gaussian
    & Negligible & High & 1 & Yes & 0.68 & 1.00 \\
\bottomrule
\end{tabular}
\end{table}

Table~\ref{tab:risk} reports two risk values. The column $R$ applies the risk
matrix of ISO/SAE 21434 Table~H.8 to the impact and feasibility ratings as the
standard specifies, and detection plays no part in it. It assigns the maximum
value of $5$ to the model-free stealthy attack and the same value of $5$ to
fuzzing, replay, plateau and denial of service, because all five combine a
Severe impact rating with a High feasibility rating. The unmodified arithmetic
therefore does not separate an attack that the deployed detectors stop from one
that they do not, and the five factors of Table~G.6 contain no term in which
that difference can be expressed.

The column $R_{\det}$ admits the measured detector response. It applies the risk formula of ISO/SAE 21434 Table~H.10, $R=1+I\cdot F$, with an evasion factor inserted,
\begin{equation}
  R_{\det} = 1 + I\cdot F\cdot E ,
  \qquad
  E = 1-\max_{j\in\mathcal{M}}\hat{A}_j ,
  \label{eq:rdet}
\end{equation}
where $I$ and $F$ are the numerical impact and feasibility values of Table~H.10
and $E$ is the measured fraction of attacked samples that the best monitor in
the battery does not flag. An attack the battery reports at a high rate is
discounted in proportion to how much of it the battery sees, and an attack the
battery does not report is not discounted at all. 
The conventional attacks occupy the lowest positions and the model-free stealthy attack the highest. Equation~\eqref{eq:rdet} is an extension proposed here and is not part of the standard. 

Of the attacks evaluated, exactly two satisfy both conditions of~\eqref{eq:eps-alpha}: the model-free construction, rated High feasibility, and the model-based construction, rated Low. Denial of service, fuzzing, replay, plateau and Gaussian injection all require less attack potential than either, and none of them is a member of that set, because the first four are reported by at least one monitor and the fifth produces no departure. Within the set of attacks that attain a safety-relevant lane departure while holding every deployed monitor at its nominal false-alarm rate, the model-free construction therefore has the highest attack feasibility rating. That set is the one an OEM must design against, and the least-capable adversary in it needs nothing beyond recorded bus traffic.


\section{Conclusion and Future Work}
\label{sec:conclusion}

This paper demonstrated that safety-critical stealthy attacks on in-vehicle networks can be constructed without privileged knowledge of vehicle dynamics, estimator parameters, or intrusion detection mechanisms. By learning nominal measurement relationships through a data-driven VARX model and calibrating a residual envelope from passively recorded traffic, the proposed adversary generated measurement-consistent false-data injections that degraded closed-loop vehicle behavior while holding every evaluated monitor within 1.4 percentage points of its nominal false-alarm rate. These findings highlight a distinction between conventional attack-feasibility assessment and the observed detectability of an attack. The evaluated impact-and-feasibility-based risk scoring assigned identical risk values to attack scenarios that exhibited substantially different detection responses. The proposed detection-conditioned risk formulation provides a complementary means of incorporating measured monitoring effectiveness into cybersecurity risk evaluation.

Future work will translate these insights into resource-efficient protection mechanisms for legacy in-vehicle networks, where bandwidth limitations, computational constraints, and stringent real-time requirements can restrict continuous cryptographic protection. The objective is to develop adaptive, resource-aware security strategies that prioritize protection where it is most effective, enabling practical and scalable defenses within the operational constraints of existing automotive communication architectures.

\renewcommand{\bibfont}{\scriptsize}
\printbibliography

\end{document}